\documentclass[sigconf, nonacm]{acmart}

\AtBeginDocument{%
  }

\setcopyright{acmcopyright}
\copyrightyear{2018}
\acmYear{2018}
\acmDOI{XXXXXXX.XXXXXXX}

\acmConference[Conference acronym 'XX]{Make sure to enter the correct
  conference title from your rights confirmation emai}{June 03--05,
  2018}{Woodstock, NY}
\acmPrice{15.00}
\acmISBN{978-1-4503-XXXX-X/18/06}

\begin{document}

\title{Generalized Posterior Calibration via Sequential Monte Carlo Sampler}

\author{Masahiro Tanaka}
\email{gspddlnit45@toki.waseda.jp}
\orcid{0000-0002-2269-8468}
\affiliation{%
  \institution{Faculty of Economics, Fukuoka University,}
  \streetaddress{8-19-1, Nanakuma, Jonan}
  \city{Fukuoka}
  \state{Fukuoka}
  \country{Japan}
  \postcode{814-0180}
}


\begin{abstract}
As the amount and complexity of available data increases, the need
for robust statistical learning becomes more pressing. To enhance
resilience against model misspecification, the generalized posterior
inference method adjusts the likelihood term by exponentiating it
with a learning rate, thereby fine-tuning the dispersion of the posterior
distribution. This study proposes a computationally efficient strategy
for selecting an appropriate learning rate. The proposed approach
builds upon the generalized posterior calibration (GPC) algorithm,
which is designed to select a learning rate that ensures nominal frequentist
coverage. This algorithm, which evaluates the coverage probability
using bootstrap samples, has high computational costs because of the
repeated posterior simulations needed for bootstrap samples. To address
this limitation, the study proposes an algorithm that combines elements
of the GPC algorithm with the sequential Monte Carlo (SMC) sampler.
By leveraging the similarity between the learning rate in generalized
posterior inference and the inverse temperature in SMC sampling, the
proposed algorithm efficiently calibrates the posterior distribution
with a reduced computational cost. For demonstration, the proposed
algorithm was applied to several statistical learning models and shown
to be significantly faster than the original GPC.
\end{abstract}



\maketitle

\section{Introduction}

As the amount of available data and the complexity of workhorse models
increase, the need for robustness in statistical learning becomes
more pressing. Given the trend, this paper focuses on two strands
of robust Bayesian methods. The first strand is generalized Bayesian
inference \citep{Martin2022,Nottforthcoming}. Let $\mathcal{D}$
denote a dataset with a sample size $N$ and consider it to be independent,
$\mathcal{D}=\left\{ \mathcal{D}_{i}\right\} _{i=1}^{N}$. In standard
Bayesian inference, the posterior of a $K$-dimensional unknown parameter
vector $\boldsymbol{\theta}$ is composed of a likelihood $p\left(\mathcal{D}|\boldsymbol{\theta}\right)$
and prior $p\left(\boldsymbol{\theta}\right)$,
\[
\pi\left(\boldsymbol{\theta}\right)\propto p\left(\mathcal{D}|\boldsymbol{\theta}\right)p\left(\boldsymbol{\theta}\right),
\]
In contrast, the generalized Bayesian inference is based on a generalized
posterior that is obtained by equipping a learning rate $\eta\left(>0\right)$
(also called a scaling parameter) with the following likelihood:
\[
\pi_{\eta}^{*}\left(\boldsymbol{\theta}\right)\propto p\left(\mathcal{D}|\boldsymbol{\theta}\right)p\left(\boldsymbol{\theta}\right)
\]
By setting $\eta<1$, we can increase the spread of the posterior,
making the inference robust to model misspecifications (e.g., \citep{Walker2001,Gruenwald2012,Miller2019,Miller2021}).
The second class of robust Bayesian methods is Gibbs posterior inference
\citep{Zhang2006a,Zhang2006b,Jiang2008,Bissiri2016,Atchade2017,Syring2023}.
This approach formulates an inferential problem using a generic loss
function $r_{i}\left(\boldsymbol{\theta};\mathcal{D}_{i}\right)$
instead of a probabilistic model:
\[
\pi_{\eta}^{*}\left(\boldsymbol{\theta}\right)\propto\exp\left\{ -Nr\left(\boldsymbol{\theta};\mathcal{D}\right)\right\} ^{\eta}p\left(\boldsymbol{\theta}\right),
\]
where $r\left(\boldsymbol{\theta};\mathcal{D}\right)$ denotes an
empirical risk function defined as 
\[
r\left(\boldsymbol{\theta};\mathcal{D}\right)=\frac{1}{N}\sum_{i=1}^{N}r_{i}\left(\boldsymbol{\theta};\mathcal{D}_{i}\right).
\]
The algorithm proposed in the paper can be applied to both Gibbs and
generalized posterior inferences. Subsequently, $q\left(\boldsymbol{\theta};\mathcal{D}\right)$
denotes a likelihood $p\left(\mathcal{D}|\boldsymbol{\theta}\right)$
or pseudo-likelihood $\exp\left\{ -Nr\left(\boldsymbol{\theta};\mathcal{D}\right)\right\} $,
\begin{equation}
\pi_{\eta}^{*}\left(\boldsymbol{\theta}\right)\propto q\left(\boldsymbol{\theta};\mathcal{D}\right)^{\eta}p\left(\boldsymbol{\theta}\right).\label{eq: pseudo-posterior}
\end{equation}

There are several approaches to select $\eta$ (\citep{Gruenwald2017,Holmes2017,Lyddon2019,Syring2019}),
and each with a different focus\footnote{See \citep{Wu2023} for a comparison.}.
We intend to improve the generalized posterior calibration (GPC) algorithm
proposed by \citet{Syring2019}. In the GPC, the coverage probability
under a specific value of $\eta$ is evaluated using bootstrap samples,
and $\eta$ is numerically chosen to achieve the nominal frequentist
coverage probability. A notable limitation of the GPC is its computational
cost. We must repeatedly run a posterior simulator for bootstrap samples
until convergence. 

The contribution of this study is that it develops a new computational
strategy for choosing the learning rate $\eta$. Essentially, the
algorithm we propose is a fusion of the GPC and sequential Monte Carlo
(SMC) samplers \citep{DelMoral2006,Dai2022}. SMC samplers are a class
of algorithms that apply importance sampling to intermediate distributions
that bridge from a prior to a posterior. The learning rate in generalized
posterior inference and the inverse temperature in SMC samplers play
the same role--powering the likelihood. By exploiting this similarity,
we transform the target distribution gradually to achieve the target
credible/confidence level. Indeed, \citet{Syring2019} mentioned a
related idea in their supplementary material.\footnote{``In our implementation of the algorithm, the posterior is sampled
every time $\omega$ {[}note by the author: learning rate{]} is updated.
However, it may be faster to sample the posterior $M$ times for $\omega_{0}$
and subsequently use importance sampling to update the posterior samples
each time $\omega$ is updated.'' (p. 2 of the Supplementary Material).} However, to the best knowledge of the author, no practical implementation
of this idea has been studied. Furthermore, it is ineffective to apply
importance sampling directly to the problem because its quality critically
depends on the disparity between the proposal and target distributions
(e.g., \citep{Liu2001}); when the consecutive learning rates, i.e.,
the proposal and target distributions, are not close enough, importance
sampling would not work well. Therefore, the use of an SMC sampler
is of critical importance.

The remainder of this paper is structured as follows. Section 2 introduces
the proposed algorithm. In Section 3, we apply the algorithm to synthetic
and real data for demonstration. Section 4 concludes the paper. 

\section{Method}

\subsection{Generalized posterior calibration}

This section describes the GPC \citep{Syring2019}. We refer to it
as the GPC-MCMC to distinguish it from the proposed algorithm. Let
$\mathcal{C}_{\alpha}^{\eta}\left(\mathcal{D}\right)$ denote the
generalized posterior $100\left(1-\alpha\right)\%$ credible set for
$\boldsymbol{\theta}$ with $\eta$. The coverage probability function
is represented as follows:
\[
c_{\alpha}\left(\eta|\mathbb{P}\right)=\mathbb{P}\left\{ \boldsymbol{\theta}^{\dagger}\left(\eta\right)\in\mathcal{C}_{\alpha}^{\eta}\left(\mathcal{D}\right)\right\} ,
\]
where $\boldsymbol{\theta}^{\dagger}\left(\eta\right)$ denotes the
Kullback-Leibler minimizer in the model obtained with $\eta$. As
the true data distribution $\mathbb{P}$ is unknown, we replace it
with the empirical distribution $\mathbb{P}_{N}$,
\[
c_{\alpha}\left(\eta|\mathbb{P}_{N}\right)=\mathbb{P}_{N}\left\{ \hat{\boldsymbol{\theta}}\left(\eta\right)\in\mathcal{C}_{\alpha}^{\eta}\left(\mathcal{D}\right)\right\} ,
\]
where $\hat{\boldsymbol{\theta}}\left(\eta\right)$ represents a point
estimate of $\boldsymbol{\theta}$ obtained with $\eta$. Even with
this modification, $c_{\alpha}\left(\eta|\mathbb{P}_{N}\right)$ cannot
be evaluated because enumeration of all $N^{N}$ possible with-replacement
samples from $\mathcal{D}$ is needed. Therefore, we approximate $\mathbb{P}_{N}$
using a bootstrap method. With $B$ bootstrap samples $\left\{ \breve{\mathcal{D}}^{\left[b\right]}\right\} _{b=1}^{B}$,
the coverage probability can be estimated as 
\[
\hat{c}_{\alpha}\left(\eta|\mathbb{P}_{N}\right)=\frac{1}{B}\sum_{b=1}^{B}\mathbb{I}\left\{ \hat{\boldsymbol{\theta}}\left(\eta\right)\in\mathcal{C}_{\alpha}^{\eta}\left(\breve{\mathcal{D}}^{\left[b\right]}\right)\right\} ,
\]
where $\mathbb{I}\left\{ \cdot\right\} $ denotes the indicator function.
$\eta$ is chosen by solving $\hat{c}_{\alpha}\left(\eta|\mathbb{P}_{N}\right)=1-\alpha$
via a stochastic approximation \citep{Robbins1951}. At the $s$th
iteration, a single step of the stochastic approximation recursion
is given as 
\begin{equation}
\eta_{\left(s+1\right)}\leftarrow\eta_{\left(s\right)}+\varsigma_{l}\left[\hat{c}_{\alpha}\left(\eta_{\left(s\right)}|\mathbb{P}_{N}\right)-\left(1-\alpha\right)\right],\label{eq: stochastic approximation}
\end{equation}
where $\left\{ \varsigma_{l}\right\} $ denotes a non-increasing sequence
such that $\sum_{l}\varsigma_{l}=\infty$ and $\sum_{l}\varsigma_{l}^{2}<\infty$.
While \citet{Syring2019} specify $\varsigma_{s}=s^{-0.51}$ in their
study, we used a variant of Keston's \citep{Kesten1958} rule: $\varsigma_{s}\left(l\right)=l^{-0.51}$
where $l$ increases by 1 only when there is a directional change
in the trajectory of $\eta_{\left(s\right)}$ and $\hat{c}_{\alpha}\left(\eta_{\left(s\right)}|\mathbb{P}_{N}\right)<1$.
This modification shortens the convergence time significantly. Let
$R$ denote the number of posterior draws used for analysis after
discarding initial draws as warmup. Figure 1 summarizes the GPC-MCMC.

\begin{figure}

\caption{Generalized posterior calibration}

\begin{raggedright}
\medskip{}
\par\end{raggedright}
\begin{raggedright}
\texttt{input:}observed dataset $\mathcal{D}$, target kernel $\pi_{\eta}^{*}\left(\cdot\right)$,
initial guess $\eta_{1}$, target credible level $\alpha$, termination
threshold $\epsilon$. 
\par\end{raggedright}
\begin{raggedright}
\medskip{}
\par\end{raggedright}
\begin{raggedright}
Generate $B$ bootstrap samples $\left\{ \breve{\mathcal{D}}^{\left[b\right]}\right\} _{b=1}^{B}$
from $\mathcal{D}$.
\par\end{raggedright}
\begin{raggedright}
Set $s\leftarrow1$ and $l\leftarrow1$.
\par\end{raggedright}
\begin{raggedright}
\texttt{while converge}
\par\end{raggedright}
\begin{raggedright}
$\quad$Compute $\hat{\boldsymbol{\theta}}$ with $\eta_{\left(s\right)}$.
\par\end{raggedright}
\begin{raggedright}
$\quad$\texttt{for} $b=1,...,B$\texttt{:}
\par\end{raggedright}
\begin{raggedright}
$\quad$$\quad$Simulate $R$ posterior draws for $\breve{\mathcal{D}}^{\left[b\right]}$
using a MCMC 
\par\end{raggedright}
\begin{raggedright}
$\quad$$\quad$$\quad$sampler with $\eta_{\left(s\right)}$.
\par\end{raggedright}
\begin{raggedright}
$\quad$$\quad$Compute the credible set $\mathcal{C}_{\alpha}^{\eta_{\left(s\right)}}\left(\breve{\mathcal{D}}^{\left[b\right]}\right)$.
\par\end{raggedright}
\begin{raggedright}
$\quad$\texttt{end for }
\par\end{raggedright}
\begin{raggedright}
$\quad$Compute the coverage probability $\hat{c}_{\alpha}\left(\eta_{\left(s\right)}|\mathbb{P}_{N}\right)$.
\par\end{raggedright}
\begin{raggedright}
$\quad$\texttt{if} $\left|\hat{c}_{\alpha}\left(\eta_{\left(s\right)}|\mathbb{P}_{N}\right)-\left(1-\alpha\right)\right|<\epsilon$\texttt{:}
\par\end{raggedright}
\begin{raggedright}
$\quad$$\quad$Set $\hat{\eta}\leftarrow\eta_{\left(s\right)}$.
\par\end{raggedright}
\begin{raggedright}
$\quad$$\quad$\texttt{break}
\par\end{raggedright}
\begin{raggedright}
$\quad$\texttt{else}
\par\end{raggedright}
\begin{raggedright}
$\quad$$\quad$Set a new learning rate $\eta_{\left(s+1\right)}$
according to (\ref{eq: stochastic approximation}).
\par\end{raggedright}
\begin{raggedright}
$\quad$$\quad$Set $s\leftarrow s+1$.
\par\end{raggedright}
\begin{raggedright}
$\quad$\texttt{end if}
\par\end{raggedright}
\begin{raggedright}
\texttt{end while}
\par\end{raggedright}
\begin{raggedright}
\texttt{return:} $\hat{\eta}$
\par\end{raggedright}
\end{figure}

\subsection{Sequential Monte Carlo sampler}

SMC samplers \citep{DelMoral2006} are a class of Monte Carlo simulation
algorithms which repeatedly apply importance sampling to a sequence
of synthetic intermediate distributions $\left\{ \pi_{t}\right\} _{t=0}^{T}$
to obtain collections of weighted particles $\left\{ w_{t}^{\left[m\right]},\boldsymbol{\theta}_{t}^{\left[m\right]}\right\} _{m=1}^{M}$.
The initial distribution $\pi_{0}$ is the prior, $\pi_{0}\left(\boldsymbol{\theta}\right)=p\left(\boldsymbol{\theta}\right)$,
and the terminal distribution $\pi_{T}$ is the target distribution,
that is, the posterior, $\pi_{T}\left(\boldsymbol{\theta}\right)=p\left(\mathcal{D}|\boldsymbol{\theta}\right)p\left(\boldsymbol{\theta}\right)$.
An intermediate distribution is specified as a likelihood-tempered
posterior:
\begin{equation}
\pi_{t}\left(\boldsymbol{\theta}\right)\propto p\left(\mathcal{D}|\boldsymbol{\theta}\right)^{\phi_{t}}p\left(\boldsymbol{\theta}\right),\label{eq: likelihood-tempered posterior}
\end{equation}
where $\left\{ \phi_{t}\right\} _{t=1}^{T}$ denotes an increasing
sequence with $\phi_{0}=0$ and $\phi_{T}=1$. $\phi_{t}$ can be
interpreted as inverse temperature. $p$ $\left\{ \tilde{\pi}_{t}\right\} _{t=0}^{T}$
denotes a sequence of auxiliary distributions:
\[
\tilde{\pi}_{t}\left(\boldsymbol{\theta}_{0:t}\right)=\pi_{t}\left(\boldsymbol{\theta}_{t}\right)\prod_{s=0}^{t-1}\mathcal{L}_{s}\left(\boldsymbol{\theta}_{s+1},\boldsymbol{\theta}_{s}\right),
\]
where $\mathcal{L}_{t}\left(\cdot,\cdot\right)$ denotes a Markov
kernel which is also called a backward kernel, where it moves back
from $\boldsymbol{\theta}_{t+1}$ to $\boldsymbol{\theta}_{t}$. $\tilde{\pi}_{t}$
is approximated using a system of weighted particles, $\left\{ w_{t}^{\left[m\right]},\boldsymbol{\theta}_{t}^{\left[m\right]}\right\} _{m=1}^{M}$.
The particles are mutated via a Markov kernel $\mathcal{K}_{t}\left(\cdot,\cdot\right)$
which is termed a forward kernel. Let $\gamma_{t}\left(\boldsymbol{\theta}\right)$
denote the unnormalized posterior density with $\phi_{t}$
\[
\gamma_{t}\left(\boldsymbol{\theta}\right)=p\left(\mathcal{D}|\boldsymbol{\theta}\right)^{\phi_{t}}p\left(\boldsymbol{\theta}\right).
\]
 The unnormalized weights are represented as
\begin{eqnarray*}
W_{t}^{\left[m\right]} & \propto & \frac{\tilde{\pi}_{t}\left(\boldsymbol{\theta}_{t}^{\left[m\right]}\right)\prod_{s=1}^{t-1}\mathcal{L}_{s}\left(\boldsymbol{\theta}_{s+1}^{\left[m\right]},\boldsymbol{\theta}_{s}^{\left[m\right]}\right)}{\zeta_{1}\left(\boldsymbol{\theta}_{1}^{\left[m\right]}\right)\prod_{s^{\prime}=1}^{t-1}\mathcal{K}_{s^{\prime}}\left(\boldsymbol{\theta}_{s^{\prime}-1}^{\left[m\right]},\boldsymbol{\theta}_{s^{\prime}}^{\left[m\right]}\right)}\\
 & \propto & \widetilde{W}_{t}^{\left[m\right]}W_{t-1}^{\left[m\right]},
\end{eqnarray*}
where $\widetilde{W}_{t}^{\left[m\right]}$ represents the unnormalized
incremental weight given by
\[
\widetilde{W}_{t}^{\left[m\right]}=\frac{\gamma_{t}\left(\boldsymbol{\theta}_{t}^{\left[m\right]}\right)\mathcal{L}_{t-1}\left(\boldsymbol{\theta}_{t}^{\left[m\right]},\boldsymbol{\theta}_{t-1}^{\left[m\right]}\right)}{\gamma_{t-1}\left(\boldsymbol{\theta}_{t-1}^{\left[m\right]}\right)\mathcal{K}_{t}\left(\boldsymbol{\theta}_{t-1}^{\left[m\right]},\boldsymbol{\theta}_{t}^{\left[m\right]}\right)}.
\]

Following the literature, we employ an MCMC kernel for the forward
kernel $\mathcal{K}_{t}$, leaving it $\pi_{t}$-invariant. This choice
is optimal, as the variance of the unnormalized weights is approximately
minimized \citep{DelMoral2006}. The backward kernel is represented
as
\[
\mathcal{L}_{t-1}\left(\boldsymbol{\theta}_{t}^{\left[m\right]},\boldsymbol{\theta}_{t-1}^{\left[m\right]}\right)=\frac{\pi_{t}\left(\boldsymbol{\theta}_{t-1}^{\left[m\right]}\right)\mathcal{K}_{t}\left(\boldsymbol{\theta}_{t-1}^{\left[m\right]},\boldsymbol{\theta}_{t}^{\left[m\right]}\right)}{\pi_{t}\left(\boldsymbol{\theta}_{t}^{\left[m\right]}\right)}.
\]
 Under this specification, the unnormalized incremental weights are
reduced to 

\[
\widetilde{W}_{t}^{\left[m\right]}=\frac{\gamma_{t}\left(\boldsymbol{\theta}_{t-1}^{\left[m\right]}\right)}{\gamma_{t-1}\left(\boldsymbol{\theta}_{t-1}^{\left[m\right]}\right)}=p\left(\mathcal{D}|\boldsymbol{\theta}_{t-1}^{\left[m\right]}\right)^{\phi_{t}-\phi_{t-1}},
\]
and thus the unnormalized weights are updated as 
\[
W_{t}^{\left[m\right]}=w_{t-1}^{\left[m\right]}q\left(\boldsymbol{\theta}_{t-1}^{\left[m\right]};\mathcal{D}\right){}^{\phi_{t}-\phi_{t-1}}.
\]

As $\pi_{t}$ evolves, the variance of the weights tends to increase,
inducing the weighted particles to degenerate. The standard metric
of particle degeneracy is the effective sample size (ESS) \citep{Kong1994},
\[
ESS_{t}=\frac{1}{\sum_{m=1}^{M}\left(w_{t}^{\left[m\right]}\right)^{2}}=\frac{\left(\sum_{m=1}^{M}W_{t-1}^{\left[m\right]}\widetilde{W}_{t}^{\left[m\right]}\right)^{2}}{\sum_{m=1}^{M}\left(W_{t-1}^{\left[m\right]}\widetilde{W}_{t}^{\left[m\right]}\right)^{2}}.
\]
This study used stratified resampling algorithm \citep{Kitagawa1996}
if $ESS_{t}$ is below a prespecified threshold $\overline{ESS}=\psi M$,
$\psi\in\left(0,1\right)$. 

The adaptive SMC sampler is applicable to generalized posterior. Exploiting
the similarity between pseudo-posterior (\ref{eq: pseudo-posterior})
in generalized/Gibbs posterior inference and likelihood tempered posterior
(\ref{eq: likelihood-tempered posterior}) in SMC sampling, we treat
the learning rate $\eta_{t}$ analogously to the inverse temperature
$\phi_{t}$ as

\[
\gamma_{t}^{*}\left(\boldsymbol{\theta}\right)=q\left(\boldsymbol{\theta};\mathcal{D}\right)^{\eta_{t}}p\left(\boldsymbol{\theta}\right).
\]

The temperature schedule is critically important in the SMC sampler.
If the temperature schedule is too coarse, the consecutive intermediate
distributions are too different to obtain a good particle approximation,
while if it is too fine, it incurs unnecessary computational load.
In this study, we use the adaptive SMC sampler \citep{Jasra2011,Beskos2016,Whiteley2016,Huggins2019}.
$\eta_{t+1}$ is chosen by keeping the ESS above a target level, $\widetilde{ESS}_{t}=\xi ESS_{t-1}$,
$\xi\in\left(0,1\right)$. The ESS can be represented as a function
of $\eta_{t}$ as follows: 
\[
ESS\left(\eta_{t}\right)=\frac{\left(\sum_{m=1}^{M}W_{t-1}^{\left[m\right]}q\left(\boldsymbol{\theta}_{t-1}^{\left[m\right]};\mathcal{D}\right)^{\eta_{t}-\eta_{t-1}}\right)^{2}}{\sum_{m^{\prime}=1}^{M}\left(W_{t-1}^{\left[m^{\prime}\right]}q\left(\boldsymbol{\theta}_{t-1}^{\left[m^{\prime}\right]};\mathcal{D}\right)\right)^{2}}.
\]
$\phi_{t}$ is chosen by solving $ESS\left(\eta_{t}\right)=\widetilde{ESS}_{t}$
via bisection search. The choice of $\xi$ involves a trade-off between
sampling quality and computational cost. The weighted particles are
resampled if $ESS\left(\phi_{1}\right)<\overline{ESS}$. Figure
2 summarizes the adaptive SMC sampler for the generalized posterior.
Let $\textrm{ASMC}^{*}\left(\eta_{0},\eta_{T};\left\{ w_{0}^{\left[m\right]},\boldsymbol{\theta}_{0}^{\left[m\right]}\right\} _{m=1}^{M}\right)$
denote the adaptive SMC sampler for the generalized posterior with
initial and terminal learning rates $\eta_{0},\eta_{T}$, respectively,
and initial weighted particles $\left\{ w_{0}^{\left[m\right]},\boldsymbol{\theta}_{0}^{\left[m\right]}\right\} _{m=1}^{M}$. 

\begin{figure}

\caption{Adaptive SMC sampler for generalized/Gibbs posterior}

\begin{raggedright}
\medskip{}
\texttt{input:} dataset $\mathcal{D}$, target kernel $\pi_{\eta}^{*}\left(\cdot\right)$,
initial and terminal learning rate $\eta_{1}$, $\eta$, target ESS
$\widetilde{ESS}$, resampling threshold $\overline{ESS}$, initial
weighted particles $\left\{ w_{0}^{\left[m\right]},\boldsymbol{\theta}_{0}^{\left[m\right]}\right\} _{m=1}^{M}$.
\par\end{raggedright}
\begin{raggedright}
\medskip{}
\par\end{raggedright}
\begin{raggedright}
Set $t\leftarrow0$.
\par\end{raggedright}
\begin{raggedright}
\texttt{while} \texttt{converge:}
\par\end{raggedright}
\begin{raggedright}
$\quad$Set $t\leftarrow t+1$
\par\end{raggedright}
\begin{raggedright}
$\quad$\texttt{if} $ESS\left(\eta\right)>\widetilde{ESS}$ \texttt{then:}
\par\end{raggedright}
\begin{raggedright}
$\quad$$\quad$Set $T\leftarrow t$ and $\eta_{T}\leftarrow\eta$
and 
\par\end{raggedright}
\begin{raggedright}
$\quad$\texttt{else:}
\par\end{raggedright}
\begin{raggedright}
$\quad$$\quad$ Solve $ESS\left(\eta_{t}\right)=\widetilde{ESS}$
in $\eta_{t}\in\left(\eta_{t-1},\eta\right]$.
\par\end{raggedright}
\begin{raggedright}
$\quad$\texttt{end if}
\par\end{raggedright}
\begin{raggedright}
$\quad$\texttt{for} $m=1,...,M$\texttt{:}
\par\end{raggedright}
\begin{raggedright}
$\quad$$\quad$Compute the unnormalized weights 
\[
W_{t}^{\left[m\right]}=w_{t-1}^{\left[m\right]}q\left(\boldsymbol{\theta}_{t-1}^{\left[m\right]};\mathcal{D}\right)^{\eta_{t}-\eta_{t-1}}.
\]
\par\end{raggedright}
\begin{raggedright}
$\quad$$\quad$Normalize the weights 
\[
w_{t}^{\left[m\right]}=W_{t}^{\left[m\right]}\left(\sum_{m^{\prime}=1}^{M}W_{t}^{\left[m^{\prime}\right]}\right)^{-1}.
\]
\par\end{raggedright}
\begin{raggedright}
$\quad$\texttt{end for}
\par\end{raggedright}
\begin{raggedright}
$\quad$Resample the particles if $ESS_{t}<\overline{ESS}$.
\par\end{raggedright}
\begin{raggedright}
$\quad$\texttt{for} $m=1,...,M$\texttt{:}
\par\end{raggedright}
\begin{raggedright}
$\quad$$\quad$Sample the particles $\boldsymbol{\theta}_{t}^{\left[m\right]}\sim\mathcal{K}_{t}\left(\boldsymbol{\theta}_{t-1}^{\left[m\right]},\cdot\right)$.
\par\end{raggedright}
\begin{raggedright}
$\quad$\texttt{end for}
\par\end{raggedright}
\begin{raggedright}
$\quad$\texttt{if} $\eta_{t}=\eta$\texttt{ then:}
\par\end{raggedright}
\begin{raggedright}
$\quad$$\quad$\texttt{break}
\par\end{raggedright}
\begin{raggedright}
$\quad$\texttt{end if}
\par\end{raggedright}
\begin{raggedright}
\texttt{end while}
\par\end{raggedright}
\begin{raggedright}
\texttt{return} $\left\{ w_{T}^{\left[m\right]},\boldsymbol{\theta}_{T}^{\left[m\right]}\right\} _{m=1}^{M}$
\par\end{raggedright}
\end{figure}

\subsection{Generalized posterior calibration via sequential Monte Carlo sampler}

This paper proposes the GPC via SMC sampler (GPC-SMC) (Figure 3),
which, as its name suggests, is a fusion of the two algorithms. Instead
of an MCMC sampler, we apply the adaptive SMC sampler to bootstrap
samples. The weighted particles are re-used as the initial states
for the next iteration. At the $s$th iteration, the adaptive SMC
sampler runs with a temperature schedule $\left\{ \eta_{\left(s\right)},...,\eta_{\left(s+1\right)}\right\} $
and initial weighted particles $\left\{ w_{\left(s\right)}^{\left[b,m\right]},\boldsymbol{\theta}_{\left(s\right)}^{\left[b,m\right]}\right\} _{m=1}^{M}$.
As the iterations proceed, a temperature schedule is likely to become
shorter, drastically reducing the computational cost per iteration.
In contrast, the GPC-MCMC has a constant computational cost per iteration;
even when $\eta_{\left(s\right)}$ and $\eta_{\left(s+1\right)}$
are close during the last phase of the optimization, it needs to simulate
the full length of chains. 

Though we can initialize particles using an SMC sampler with a long
temperature schedule $\left\{ 0,...,\eta_{\left(1\right)}\right\} $,
we suggest using an MCMC sampler with $\eta_{\left(1\right)}$ to
generate initial particles. The corresponding weights are set according
to the posterior densities evaluated at the particles. 
\begin{eqnarray*}
W_{\left(1\right)}^{\left[b,m\right]} & = & q\left(\boldsymbol{\theta}^{\left[b,m\right]};\breve{\mathcal{D}}^{\left[b\right]}\right)^{\eta_{\left(1\right)}}p\left(\boldsymbol{\theta}^{\left[b,m\right]}\right),\\
w_{\left(1\right)}^{\left[b,m\right]} & = & W_{\left(1\right)}^{\left[b,m\right]}\left(\sum_{m^{\prime}=1}^{M}W_{\left(1\right)}^{\left[b,m^{\prime}\right]}\right)^{-1}.
\end{eqnarray*}

In the GPC-SMC, a temperature schedule can be decreasing if $\eta_{\left(s\right)}<\eta_{\left(s+1\right)}$.
This might seem peculiar, but the SMC sampler effectively works as
long as the temperature schedule is selected finely enough, that is,
$\xi$ is close to one, making the proposal and target distributions
sufficiently similar. 

\begin{figure}

\caption{Generalized posterior calibration via SMC sampler}

\begin{raggedright}
\medskip{}
\par\end{raggedright}
\begin{raggedright}
\texttt{input:} observed dataset $\mathcal{D}$, target kernel $\pi_{\eta}^{*}\left(\cdot\right)$,
initial guess $\eta_{\left(1\right)}$, target credible level $\alpha$,
termination threshold $\epsilon$.
\par\end{raggedright}
\begin{raggedright}
\medskip{}
\par\end{raggedright}
\begin{raggedright}
Set $s\leftarrow1$ and $l\leftarrow1$.
\par\end{raggedright}
\begin{raggedright}
Compute the posterior estimate $\hat{\boldsymbol{\theta}}$ with $\eta_{\left(1\right)}$.
\par\end{raggedright}
\begin{raggedright}
Generate $B$ bootstrap samples $\left\{ \breve{\mathcal{D}}^{\left[b\right]}\right\} _{b=1}^{B}$
from $\mathcal{D}$.
\par\end{raggedright}
\begin{raggedright}
\texttt{for} $b=1,...,B$\texttt{:}
\par\end{raggedright}
\begin{raggedright}
$\quad$Simulate $M$ posterior draws $\left\{ \boldsymbol{\theta}_{\left(1\right)}^{\left[b,m\right]}\right\} _{m=1}^{M}$
using a MCMC
\par\end{raggedright}
\begin{raggedright}
$\quad$$\quad$sampler with $\eta_{\left(1\right)}$.
\par\end{raggedright}
\begin{raggedright}
$\quad$Initialize the weights $\left\{ w_{\left(1\right)}^{\left[b,m\right]}\right\} _{m=1}^{M}$.
\par\end{raggedright}
\begin{raggedright}
$\quad$Compute the credible set $\mathcal{C}_{\alpha}^{\eta_{\left(1\right)}}\left(\breve{\mathcal{D}}^{\left[b\right]}\right)$.
\par\end{raggedright}
\begin{raggedright}
\texttt{end for}
\par\end{raggedright}
\begin{raggedright}
\texttt{while converge:}
\par\end{raggedright}
\begin{raggedright}
$\quad$\texttt{if} $\left|\hat{c}_{\alpha}\left(\eta_{\left(s\right)}|\mathbb{P}_{N}\right)-\left(1-\alpha\right)\right|<\epsilon$\texttt{
then:}
\par\end{raggedright}
\begin{raggedright}
$\quad$$\quad$$\quad$Set $\hat{\eta}\leftarrow\eta_{\left(s\right)}$.
\par\end{raggedright}
\begin{raggedright}
$\quad$$\quad$$\quad$\texttt{break}
\par\end{raggedright}
\begin{raggedright}
$\quad$\texttt{else:}
\par\end{raggedright}
\begin{raggedright}
$\quad$$\quad$Set a new learning rate $\eta_{\left(s+1\right)}$
according to (\ref{eq: stochastic approximation}).
\par\end{raggedright}
\begin{raggedright}
$\quad$$\quad$Compute the posterior estimate $\hat{\boldsymbol{\theta}}$with
$\eta_{\left(s+1\right)}$.
\par\end{raggedright}
\begin{raggedright}
$\quad$$\quad$\texttt{for} $b=1,...,B$\texttt{:}
\par\end{raggedright}
\begin{raggedright}
$\quad$$\quad$$\quad$Simulate the posterior draws
\par\end{raggedright}
\begin{raggedright}
$\quad$$\quad$$\quad$$\quad$$\left\{ w_{\left(s+1\right)}^{\left[b,m\right]},\boldsymbol{\theta}_{\left(s+1\right)}^{\left[b,m\right]}\right\} _{m=1}^{M}$
using
\par\end{raggedright}
\begin{raggedright}
$\quad$$\quad$$\quad$$\quad$ $\textrm{ASMC}^{*}\left(\eta_{\left(s\right)},\eta_{\left(s+1\right)};\left\{ w_{\left(s\right)}^{\left[b,m\right]},\boldsymbol{\theta}_{\left(s\right)}^{\left[b,m\right]}\right\} _{m=1}^{M}\right)$.
\par\end{raggedright}
\begin{raggedright}
$\quad$$\quad$$\quad$Compute the credible set $\mathcal{C}_{\alpha}^{\eta_{\left(s+1\right)}}\left(\breve{\mathcal{D}}^{\left[b\right]}\right)$.
\par\end{raggedright}
\begin{raggedright}
$\quad$$\quad$\texttt{end for}
\par\end{raggedright}
\begin{raggedright}
$\quad$$\quad$Set $s\leftarrow s+1$.
\par\end{raggedright}
\begin{raggedright}
$\quad$\texttt{end if}
\par\end{raggedright}
\begin{raggedright}
\texttt{end while}
\par\end{raggedright}
\begin{raggedright}
\texttt{return:} $\hat{\eta}$
\par\end{raggedright}
\end{figure}

In our context, each of an MCMC sampler and SMC sampler has its advantages
and disadvantages. The computational cost of an SMC sampler is generally
larger than that of a MCMC sampler. Therefore, if an efficient MCMC
sampler is available, there may be no strong reason to use an SMC
sampler. On the contrary, an SMC sampler works well for distributions
that are difficult to simulate, e.g., posterior distributions with
multimodality or discontinuity. The advantage of an SMC sampler is
more pronounced for Gibbs posterior inference because no closed-form
expression of the conditional posterior distributions is available,
making posterior simulations difficult, and because for many cases
a Gibbs posterior is discontinuous as in the subsequent section. While
Hamiltonian Monte Carlo is a state-of-the-art solution to sampling
from arbitrary distributions, it is only applicable to continuous
target distributions. Thus, as seen below, there is no feasible option
for posterior simulation except classical and sub-optimal algorithms
such as the random-walk Metropolis-Hastings algorithm. Indeed, the
choice of an MCMC kernel $\mathcal{K}_{t}$ is also important in SMC
sampling, but it is not as sensitive an issue as in MCMC sampling. 

\section{Application}

\subsection{Quantile regression}

For demonstration, we applied the GPC-SMC and GPC-MCMC to quantile
regression with synthetic data. Following Section 4 of \citep{Syring2019},
the DGP is specified as 
\[
y_{i}=\boldsymbol{\theta}^{\top}\boldsymbol{x}_{i}+\varepsilon_{i},\quad\varepsilon_{i}\sim\mathcal{N}\left(0,\sigma^{2}\right),
\]
where $\boldsymbol{x}_{i}=\left(1,x_{1,i}\right)^{\top}$, $x_{1,i}+2\sim\chi_{4}^{2}$,
$\varepsilon_{i}\sim\mathcal{N}\left(0,1\right)$, $\boldsymbol{\theta}=\left(\theta_{1},\theta_{2}\right)^{\top}=\left(2,1\right)^{\top}$,
and $\sigma^{2}=1$. We consider three cases of the sample size $N\in\left\{ 100,400,1600\right\} $.
We infer $\boldsymbol{\theta}$ using a Gibbs posterior approach.
$\boldsymbol{\theta}$ is inferred for the $\left(100\times\tau\right)$-th
percentile based on an empirical risk function given by
\[
r\left(\boldsymbol{\theta};\mathcal{D}\right)=\frac{1}{N}\sum_{i=1}^{N}\left|\left(y_{i}-\boldsymbol{\theta}^{\top}\boldsymbol{x}_{i}\right)\left(\tau-\mathbb{I}\left\{ y_{i}<\boldsymbol{\theta}^{\top}\boldsymbol{x}_{i}\right\} \right)\right|.
\]
We fix $\tau=0.5$. The prior for $\boldsymbol{\theta}$ is a normal
distribution with mean zero and variance $\varsigma^{2}$, $\boldsymbol{\theta}\sim\mathcal{N}\left(\boldsymbol{0}_{2},\varsigma^{2}\boldsymbol{I}_{2}\right)$.
We choose a fairly diffuse prior, $\varsigma^{2}=100^{2}$. 

For the GPC-SMC, we employed the adaptive random-walk Metropolis-Hastings
(RWMH) algorithm \citep{Haario2001} as the MCMC kernel. At the $t$th
iteration of SMC sampling, a proposal $\boldsymbol{\theta}^{\prime}$
is generated from a multivariate normal distribution, $\boldsymbol{\theta}^{\prime}\sim\mathcal{N}\left(\boldsymbol{\theta}_{t}^{\left[m\right]},\zeta_{t}\boldsymbol{\Sigma}_{t}\right)$,
where $\zeta_{t}\left(>0\right)$ denotes a scaling parameter and
$\boldsymbol{\Sigma}_{t-1}$ represents a covariance matrix. $\boldsymbol{\theta}^{\prime}$
is accepted with probability: 
\[
\alpha\left(\boldsymbol{\theta}^{\prime},\boldsymbol{\theta}_{t-1}^{\left[m\right]}\right)=\frac{\gamma_{t}^{*}\left(\boldsymbol{\theta}^{\prime}\right)}{\gamma_{t}^{*}\left(\boldsymbol{\theta}_{t-1}^{\left[m\right]}\right)}.
\]
$\boldsymbol{\Sigma}_{t}$ is chosen based on the covariance matrix
of the current particles as
\[
\boldsymbol{\Sigma}_{t+1}\leftarrow\frac{\eta_{t}}{\eta_{t+1}}\frac{1}{M}\sum_{m=1}^{M}w_{t}^{\left[m\right]}\boldsymbol{\theta}_{t}^{\left[m\right]}\left(\boldsymbol{\theta}_{t}^{\left[m\right]}\right)^{\top}.
\]
$\zeta_{t}$ is adaptively tuned on the fly according to the following
updating rule:
\[
\log\zeta_{t+1}\leftarrow\log\zeta_{t}+\varphi_{t}\left(\varrho^{*}-\varrho_{t}\right),
\]
where $\varrho_{t}$ represents the average acceptance rate at the
$t$th iteration, $\varrho^{*}$ denotes a prespecified target acceptance
rate, and $\left\{ \varphi_{t}\right\} $ indicates a decreasing sequence
such that $\sum_{t}\varphi_{t}=\infty$ and $\sum_{t}\varphi_{t}^{2}<\infty$.
We fixed $\varrho^{*}=0.25$ and $\varphi_{t}=\left(t+1\right)^{-0.51}$.

We evaluated the numerical efficiency of the two algorithms based
on wall clock computation time. It is impossible to exactly compare
the two algorithms, because the relative performance critically depends
on the implementation details, in particular, the number of MCMC draws
$R$ and number of particles $M$. We choose $M=1,000$ and $R=20,000$
so that the multivariate effective sample size (multiESS) \citep{Vats2019}
is approximately the same as $M$. Although multiESS for MCMC sampling,
which reflects the degrees of autocorrelations in the generated chain,
is conceptually different from ESS for SMC sampling, they have similar
implications for practitioners. As a metric of the efficiency of an
MCMC sampler, a minimum of chain-wise effective sample sizes (minESS)
\citep{Geyer1992} is also used in the literature. As minESS is generally
smaller than multiESS (roughly a half in our case), using multiESS
means that the comparison is more favorable to the GPC-MCMC than using
minESS.

The other tuning parameters are selected as follows. The target credible
level is $\alpha=0.05$, meaning a 95\% credible set is calibrated.
The number of bootstrap samples is $B=500$. The stopping criterion
for finding the appropriate $\eta$ is $\epsilon=0.005$. The initial
guess of the learning rate is fixed to $\eta=1$. The threshold for
resampling is $\psi=0.5$, following the standard practice in the
literature. The tuning parameter of the target ESS for choosing a
new learning rate is $\xi=0.999$. While there is neither consensus
nor a principled guideline regarding the choice of $\xi$, our choice
may be one of the largest values among those used in the literature
\footnote{While \citet{Jasra2011} uses $\xi=0.95$, there is considerable latitude
in setting $\xi$: from 1/3 \citep{Kantas2014} to 0.999 \citep{Zhou2016}.}.%
Thus, our choice is rather unfavorable to the GPC-SMC in comparison
to the GPC-MCMC.

All the computations are conducted using MATLAB (R2023b) on an Ubuntu
desktop (22.04.4 LTS) running on an AMD Ryzen Threadripper 3990X 2.9GHz
64-core processor. The computation using different bootstrap samples
is parallelized. If a multi-machine parallel system is available,
the GPC-SMC would run even faster, as the particle-wise computation
in an SMC sampler is also parallelizable. For the GPC-MCMC, we use
an algorithm proposed by \citet{Vihola2012} which is a variant of
\citet{Haario2001}, because the adaptive RWMH algorithm does not
work well, resulting in strongly autocorrelated chains. Vihola's \citep{Vihola2012}
adaptive algorithm is designed to estimate the shape of the target
distribution while coercing the acceptance rate. Therefore, again,
our comparison is unfavorable to the GPC-SMC.

Table 1 reports the coverage probabilities based on 200 synthetic
datasets. For all three cases, the coverage probabilities are close
to 95\%, which implies that both algorithms effectively achieved the
target credible/confidence level. The medians of the calibrated learning
rate are $\eta\approx1.6$ for $N=100$, $\eta\approx1.4$ for $N=400$,
and $\eta\approx1.1$ for $N=1600$. 

\begin{table}
\caption{Coverage probability}

\medskip{}

\begin{centering}
\begin{tabular}{rrr}
\hline 
\multicolumn{1}{c}{$N$} & \multicolumn{2}{c}{Coverage probability (\%)}\tabularnewline
\cline{2-3} \cline{3-3} 
 & GPC-MCMC & GPC-SMC\tabularnewline
\hline 
100 & 97.0 & 96.0\tabularnewline
400 & 93.5 & 94.0\tabularnewline
1600 & 95.0 & 94.5\tabularnewline
\hline 
\end{tabular}
\par\end{centering}
\medskip{}

\centering{}%
\begin{minipage}[t]{0.9\columnwidth}%
Note: $R$ is the number of draws generated by an MCMC sampler. $M$
is the number of particles generated by an SMC sampler. Coverage probability
of ground truth is evaluated based on 200 synthetic datasets.%
\end{minipage}
\end{table}

Figure 4 displays the distributions of the computation time. The GPC-SMC
is generally faster than the GPC-MCMC. When the initial value $\eta=1$
is close to the desirable value, the difference in computation time
between the two algorithms is pronounced. This happens because the
advantage of the GPC-SMC is significant for the final phase of the
stochastic approximation optimization. For cases with $N=100$, $\eta=1$
is too small and during the initial phase, long temperature schedules
are used for SMC sampling, incurring non-negligible computational
cost. Therefore, with a good guess of $\eta$, the GPC-SMC would further
overwhelm the GPC-MCMC.

\begin{figure}
\caption{Comparison of computational cost}

\medskip{}

\begin{centering}
\includegraphics[scale=0.55]{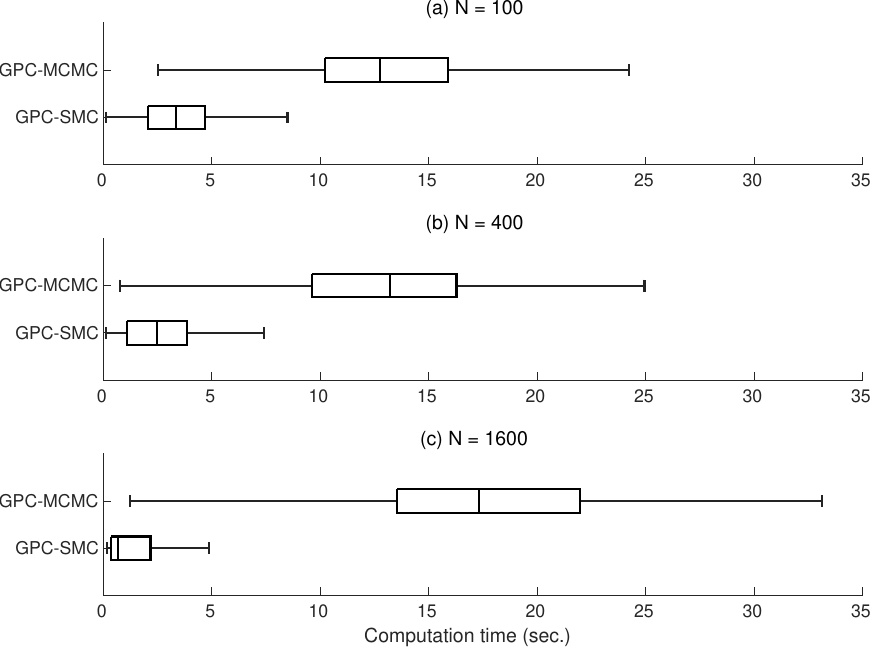}
\par\end{centering}
\medskip{}

\centering{}%
\begin{minipage}[t]{0.9\columnwidth}%
Note: Boxplot displays the distribution of computation time (in second)
for 200 synthetic datasets.%
\end{minipage}
\end{figure}

\subsection{Support vector machine}

We consider support vector machine classification with the South African
Heart Disease dataset (see Section 4.4.2 of \citep{Hastie2009}),
as in Section 5 of \citep{Syring2019}. We have a binary outcome $y_{i}\in\left\{ -1,1\right\} $
and a $K$-dimensional vector of predictors $\boldsymbol{x}_{i}=\left(1,x_{1,i},...,x_{K-1,i}\right)^{\top}$.
Stacking them yields $\boldsymbol{y}=\left(y_{1},...,y_{N}\right)^{\top}$
and $\boldsymbol{X}=\left(\boldsymbol{x}_{1},...,\boldsymbol{x}_{N}\right)^{\top}$.
The support vector machine seeks to find $\boldsymbol{\theta}=\left(\theta_{1},...,\theta_{k}\right)^{\top}$
that minimizes the following objective function:
\[
r\left(\boldsymbol{\theta}\right)=\frac{1}{N}\sum_{i=1}^{N}2\max\left(0,1-y_{i}\boldsymbol{\theta}^{\top}\boldsymbol{x}_{i}\right).
\]
We assigned an independent Laplace-type prior to $\boldsymbol{\theta}$.
The log pseudo-posterior is represented as
\[
\pi_{\eta}^{*}\left(\boldsymbol{\theta}\right)\propto-\eta\sum_{i=1}^{N}2\max\left(0,1-y_{i}\boldsymbol{x}_{i}^{\top}\boldsymbol{\theta}\right)-\nu^{-1}\sum_{k=1}^{K}\left|\frac{\theta_{k}}{\sigma_{k}}\right|,
\]
where $\sigma_{k}$ denotes the standard deviation of the $k$th predictor
$x_{k,1},...,x_{k,N}$ with $\sigma_{1}=1$ and $\nu\left(>0\right)$
represents a tuning parameter. We fixed $\nu=10$. \citep{Polson2011}
developed a Gibbs sampler for Bayesian inference of this model based
on a data augmentation technique, although a resulting credible set
($\eta=1$) is not well-calibrated \citep{Syring2019}. We use the
Gibbs sampler of \citep{Polson2011} for the GPC-MCMC, while the adaptive
RWMH for the GPC-SMC. The sample size is $N=462$ and the dimension
of $\boldsymbol{\theta}$ is $K=8$. We choose $R=20,000$ and $M=4,000$
so that the multiESS from MCMC sampling and ESS from SMC sampling
are roughly matched. The other tuning parameters are the same as in
Section 3.1.

We executed the two algorithms 20 times with different random seeds.
Both algorithms consistently yielded $\eta\approx0.09$, which aligns
with the result of \citep{Syring2019}; the Gibbs sampling algorithm
of \citep{Polson2011} without calibration of $\eta$ brings excessively
optimistic uncertainty quantification. As shown in Figure 5, the GPC-SMC
outperforms the GPC-MCMC in terms of computing speed. The medians
of computation time for the GPC-MCMC and GPC-SMC are approximately
18.0 and 10.7 minutes, respectively. In addition, the variation in
time taken for computation is much smaller for the GPC-SMC; the coefficients
of variation for the GPC-MCMC and GPC-SMC are 0.335 and 0.017, respectively. 

\begin{figure}
\caption{Comparison of computational cost}

\medskip{}

\begin{centering}
\includegraphics[scale=0.5]{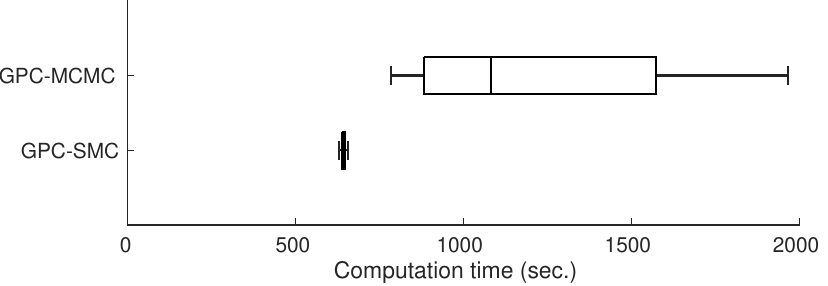}
\par\end{centering}
\medskip{}

\centering{}%
\begin{minipage}[t]{0.9\columnwidth}%
Note: Boxplot displays the distribution of computation time (in second)
for 20 runs with different random seeds.%
\end{minipage}
\end{figure}

\section{Conclusion}

In conclusion, this paper presented a novel and computationally efficient
strategy for selecting an appropriate learning rate in generalized
posterior inference. By building upon the GPC algorithm \citep{Syring2019},
which aims to achieve nominal frequentist coverage, we devised an
algorithm that combined elements of the GPC with the SMC sampler.
This integration harnessed the similarity between the learning rate
in generalized posterior inference and the inverse temperature in
SMC sampling, enabling the calibration of the posterior distribution
with reduced computational costs. The proposed approach addresses
the limitation of high computational costs associated with existing
methods by leveraging the efficiency of SMC sampling. Through empirical
demonstrations on statistical learning models, we illustrated the
effectiveness and practicality of our proposed algorithm in selecting
an appropriate learning rate for robustness toward model misspecification.

Overall, this research contributes to advancing the methodology of
generalized posterior inference by providing a scalable and efficient
solution for learning rate selection. Future work may explore further
refinements and extensions of the proposed algorithm, as well as its
application in other domains requiring robust Bayesian inference under
model uncertainty.

\bibliographystyle{ACM-Reference-Format}
\bibliography{reference}

\end{document}